\documentclass[aps,twocolumn,superscriptaddress,floatfix,nolongbibliography,noeprint]{revtex4-2}
\usepackage[colorlinks=true, citecolor=cyan, urlcolor=blue]{hyperref}
\usepackage{algorithm}
\usepackage{algpseudocode}
\usepackage{blindtext}
\usepackage{bm}
\usepackage{dsfont}
\usepackage{braket}
\usepackage{hyperref}
\usepackage{amssymb}
\usepackage{amsmath}
\usepackage{graphicx}
\usepackage{float}
\usepackage{bbold}
\usepackage{xcolor}
\usepackage{mathtools}
\newcommand{\vect}[1]{\ensuremath{\mathbf{#1}}}
\renewcommand{\t}[1]{\textrm{#1}}
\newcommand{\RNum}[1]{\uppercase\expandafter{\romannumeral #1\relax}}

\begin{document}
%\title{Universal short- and asymptotic-time sensitivity scalings in Markovian quantum metrology}

\title{Universal time scalings of sensitivity in Markovian quantum metrology}

\author{Arpan Das}
\affiliation{Faculty of Physics, University of Warsaw, Pasteura 5, 02-093 Warszawa, Poland}

\author{Wojciech G{\'o}recki}
\affiliation{INFN Sez. Pavia, via Bassi 6, I-27100 Pavia, Italy}

\author{Rafa{\l} Demkowicz-Dobrza{\'n}ski}
\affiliation{Faculty of Physics, University of Warsaw, Pasteura 5, 02-093 Warszawa, Poland}

\begin{abstract}
Assuming Markovian time evolution of a quantum sensing system, we study the general characterization of the optimal sensitivity scalings with time, under most general quantum control protocols. We allow the estimated parameter to influence both the Hamiltonian as well as the dissipative part of the quantum master equation and focus on the asymptotic-time along with the short-time sensitivity scalings. We find that via simple algebraic conditions (in terms of the Hamiltonian, the jump operators as well as their parameter derivatives), one can characterize the four classes of metrological models that represent: quadratic-linear, quadratic-quadratic, linear-linear and linear-quadratic time scalings. We also investigate the relevant time scales on which the transition between the two regimes appears. Additionally, we provide universal numerical methods to obtain quantitative bounds on sensitivity that are the tightest that exist in the literature. Simplicity and universality of our results make it suitable for diverse applications in quantum metrology.
\end{abstract}

\maketitle

\paragraph*{Introduction.}
With rapid technological advancements, need for precise characterizations of physical systems is getting increasingly demanding. Quantum metrology \cite{Giovannetti2011, Toth2014, Degen2017, Braun2018, Pezze2018, Pirandola2018}, based on the theoretical framework of quantum estimation theory \cite{Holevo, Braunstein1994} is a pursuit towards this goal. Using quantum resources \cite{resource-review} like coherence \cite{plenio-coherence} and entanglement \cite{entanglement-review} to achieve the optimal enhancement over classical strategies is crucial for further development of already spectacular achievements in   
%Most distinguished example of this is to overcome the linear scaling and obtain a quadratic (Heisenberg) scaling of achievable precision with respect to the number of resources used.
gravitational wave detection \cite{Schnabel2010, Grote2013, Ligo2019}, spectroscopy \cite{Sanders1995, Bollinger1996, Huelga1997, Leibfried2004, Roos2006}, magnetometry \cite{Jones2009, Schmitt2017, Zhibo2020} or atomic clocks \cite{Schmidt2005, Kruse2016, Hosten2016, Pezze2020, Kaubruegger2021} to name a few.

One of the pivotal moments in the development of the field, was a realization that the potential quadratic gain in the scaling of precision with the number of resources used (particles, time), referred to as the Heiseneberg scaling \cite{Holland1993, Bollinger1996, Lee2002, Giovaennetti2006, Dowling2008}, is extremely fragile in the presence of noise %, when inspecting the large-resource limit (many particles, long interrogations times) 
\cite{Huelga1997, Rubin2007, Shaji2007, Huver2008, Olivares2007, Gilbert2008, Rafal2009, Banaszek2009, Ono2010, Escher2011, Demkowicz-Dobrzanski2012}. In commonly encountered noisy scenarios, one may asymptotically expect at most a constant quantum enhancement factor, even if all possible adaptive quantum strategies are allowed \cite{Escher2011, Demkowicz-Dobrzanski2012, Demkowicz-Dobrzanski2014,  Demkowicz-Dobrzanski2017, Zhou2017, Zhou2020, Yamamoto2022}. Nevertheless, there are also cases, when noise may be efficiently countered via quantum-error correction inspired protocols and long-time or large-scale benefits of quantum coherence/entanglement may be exploited \cite{Dur2014, Kessler2014a, Arrad2014, Unden2016, Sekatski2016, Zhou2017, Zhou2017, Zhou2020}.

\begin{figure}[t]
\includegraphics[width=\columnwidth]{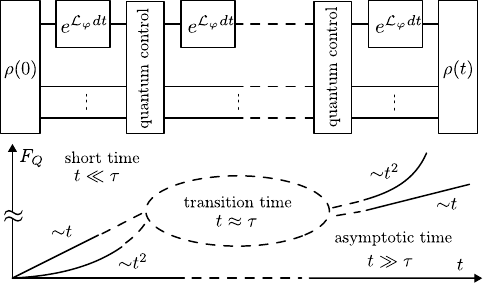}
    \caption{A general model of Markovian quantum sensing. A Quantum probe may be entangled with arbitrary large ancillary systems and  fast and strong quantum controls can be applied during the whole sensing process.
    The character of the Markovian dynamics determines the optimal quantum Fisher information scaling both in the short-time as well as in the asymptotic-time regimes and $\tau$ denotes the characteristic timescales of transition between them.  The scaling may be either linear or quadratic in both regimes, and all four combinations are allowed.}
\label{fig:intro}
\end{figure}

In this paper, we consider the most general Markovian quantum sensing model, where the dynamics of the sensing probe is governed by  the celebrated Gorini-Kossakowski-Sudarshan-Lindblad (GKSL) \cite{GKSL, lindblad76, breuer02} equation:
\begin{multline}
\label{GKSL}
    \frac{d \rho(t)}{dt}= \mathcal{L}_{\varphi}[\rho(t)] = -i[H(\varphi), \rho(t)]+ \\ \sum_{k=1}^n\left(L_{k}(\varphi)\rho(t) L_{k}^\dagger(\varphi)-\frac{1}{2}\{L_k^\dagger(\varphi) L_k(\varphi), \rho(t)\}\right),
\end{multline}
where the parameter $\varphi$ to be estimated may be encoded both in the Hamiltonian and the jump operators $L_k$. In what follows, for simplicity of notation, we will drop the explicit dependence of $H$ and $L_k$ on $\varphi$.
We assume, that we can perform arbitrary quantum control operations on the sensing probe as often as required, including entangling operations with ancilla of arbitrary size, see Fig.~\ref{fig:intro}---we implicitly assume that $H$ as well as $L_k$ operators act trivially on the ancillary system.  

We assess quantitatively the performance of metrological protocols by studying the maximal achievable quantum Fisher information (QFI),  as a function of the total sensing time $t$,  inverse of which  provides a lower bound on the achievable parameter estimation variance \cite{Helstrom1976, Braunstein1994}. There are a number of efficient numerical tools to find a quantum control protocol that yields a reasonable  sensing performance \cite{Degen2017, Zhang2024, Kurdzialek2024}, but the exact optimization of the QFI over the most general sensing strategies is typically not feasible, unless one considers idealized noiseless scenarios \cite{Giovaennetti2006}. 

The alternative approach is to analyze fundamental bounds on the performance of the most general control protocols, which are relatively easy to compute for Markovian models \cite{Fujiwara2008, Escher2011, Demkowicz-Dobrzanski2012, Demkowicz-Dobrzanski2014, Sekatski2016, Pirandola2017, Demkowicz-Dobrzanski2017, Zhou2017,  Zhou2020, Wan2022, adaptive-new-bound}. The bounds were originally developed  with parallel channel estimation schemes in mind \cite{Fujiwara2008, Escher2011, Demkowicz-Dobrzanski2012}, but were later generalized to cover the most general adaptive channel estimation schemes \cite{Demkowicz-Dobrzanski2014, Zhou2020, adaptive-new-bound} as well as continuous time Markovian sensing scenarios \cite{Demkowicz-Dobrzanski2017, Zhou2017, Wan2022}. 

These approaches allowed for an identification of simple if and only if conditions on the achievability of the asymptotic Heisenberg scaling of the QFI in Markovian Hamiltonian parameter sensing scenarios  \cite{Demkowicz-Dobrzanski2017, Zhou2017}. Bounds were shown to be asymptotically saturable \cite{Zhou2017, Zhou2020, adaptive-new-bound}, yet were not particularly tight on the short and intermediate time scales nor did they allow to obtain a proper insight into the time scales on which the transition from the short-time behaviour to the asymptotic one takes place.    
Moreover, in most of the works except e.g. \cite{chen2020, Wan2022}, the sole focus was on the parameter that appears in the Hamiltonian part of the Markovian dynamics, restricting the full generality of the studies. 

In this paper we remedy all these deficiencies. We provide a simple algebraic classification of all the four classes of Markovian sensing models, that differ by either short or asymptotic-time scaling character (linear or quadratic) as well as provide simple formulas capturing the characteristic time scales for the transitions between these regimes. Finally, we provide an efficient numerical method to compute explicit bounds that are not only tight in the short and the asymptotic-time regimes, but also perform the best in the intermediate-time regime from all known efficiently computable bounds. 
 
%  Our approach can also be anturally geenralized to  time dependent non-homogenous dynamics, like in  PengAdaptiveControTimeDependentHamiltonians2017 
\paragraph*{Scaling analysis.}
In what follows, we will use the notation, were $\vec{L}=[L_1,\dots,L_n]^T$ is a column vector of length $n$ collecting all the jump operators, while $\vect{L} = [\openone, L_1,\dots,L_n]^T$ is an extended vector with additional
identity operator at the first position. 
Our starting point is the fundamental bound on the rate of the increase of QFI derived in \cite{adaptive-new-bound}, which in the continuous time Markovian limit takes the following form (for the derivation see Supplementary Material, Appendix F in \cite{adaptive-new-bound}):   
\begin{equation}
\label{QFI-growth}
    \frac{d F_Q(t)}{dt} \leq 4\min_{h} \left(\|\mathfrak{a}(h)\|+\|\mathfrak{b}(h)\|\sqrt{F_Q(t)}\right),
\end{equation}
where $F_Q(t)$ denotes the QFI and the optimization variable
\begin{equation}
h = 
    \begin{pmatrix}
        \begin{array}{c|c}
             h_{0}^0 & \vec{h}^\dagger \\
             \hline
             \vec{h} & \mathfrak{h} \\    
        \end{array}
    \end{pmatrix}
\end{equation}
is an $n+1 \times n+1$ Hermitian matrix with a block structure, where $\mathfrak{h} \in \mathbb{C}^{n \times n}_{\t{H}}$ is an $n\times n$ hermitian matrix, $\vec{h} \in \mathbb{C}^n$ is a complex vector of length $n$, $h^0_0 \in \mathbb{R}$, $\| \cdot \|$ is the operator norm, while $\mathfrak{a}(h)$, $\mathfrak{b}(h)$ are operators defined as:
\begin{eqnarray}
\label{eq:aoperator}
\mathfrak{a}(h) & = & {\left(i\dot{\vec{L}}+ \mathfrak{h} \vec{L} + \vec{h} \openone \right)}^\dagger \left(i\dot{\vec{L}}+ \mathfrak{h} \vec{L} + \vec{h} \openone\right),\\
\label{eq:boperator}
\mathfrak{b}(h) & = &\dot{H} - \frac{i}{2} \left( \dot{\vec{L}}^\dagger \vec{L} - \vec{L}^\dagger \dot{\vec{L}} \right)  + \vect{L}^\dagger h \vect{L},
\end{eqnarray}
where $\dot{H}$, $\dot{\vec{L}}$  are derivatives of the Hamiltonian and the jump operators over the estimated parameter $\varphi$. Note that for the sake of clarity, the notation we use in the paper  is slightly modified compared with the one originally used in \cite{adaptive-new-bound}. In particular, $\mathfrak{a}(h)$, $\mathfrak{b}(h)$ represent here $\alpha^{(1)}(h)$, $-i\beta^{(1)}(h)$ respectively from \cite{adaptive-new-bound}, which are the first order time expansions of operators $\alpha$, $\beta$ commonly used in expressions for the bounds in the quantum channel estimation framework \cite{Demkowicz-Dobrzanski2012, Demkowicz-Dobrzanski2014, Zhou2020, adaptive-new-bound}.

\begin{figure*}[t]
\includegraphics[width=1.7\columnwidth]{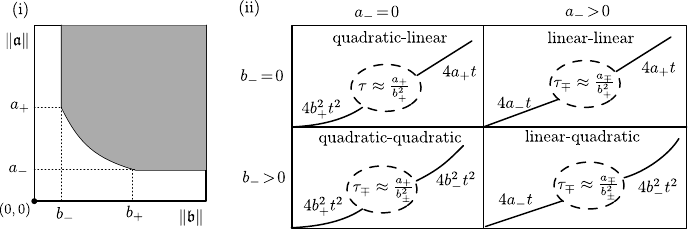}
    \caption{Schematic plot of (i) achievable pairs of operator norms $(\|\mathfrak{a}\, \|\mathfrak{b}\|)$ of operators defined in (\ref{eq:aoperator},\ref{eq:boperator}) and (ii) the short-time and asymptotic-time QFI scalings as well as the relevant transition time scales.} \label{fig:abplot}
\end{figure*}

In order to analyze both quantitative and qualitative (scalings) implications of \eqref{QFI-growth} it is best to start with the plot of the achievable region for the operator norms of $\mathfrak{a}$, $\mathfrak{b}$,   
$(\|\mathfrak{a}\|, \|\mathfrak{b}\|)$---see Fig.~\ref{fig:abplot}(i). In what follows we will refer to this plot as the `ab-plot'.
This plot can be generated as a solution of  simple semi-definite programme (SDP), see Ref. \cite{supp}, Section A. It is enough to determine whether the minimal achievable values of ${a}_{-} = \min_h\| \mathfrak{a}(h)\|$ and 
${b}_- = \min_h \|\mathfrak{b}(h)\|$ are strictly greater than zero.

If $b_-=0$, then for large enough $t$ it is always optimal to choose $\|\mathfrak{b}\|=0$ in \eqref{QFI-growth}. Consequently, the QFI will grow linearly in the asymptotic-time regime according to
the formula $F_Q \approx 4 a_+ t$, where
${a}_+ = \min_{h, \|\mathfrak{b}\| = b_-}(\|\mathfrak{a}(h)\|)$. On the other hand, if $b_->0$ the asymptotic-time QFI scaling is necessarily quadratic, $F_Q \approx 4 b_-^2 t^2$. 

In a complementary manner, $a_-=0$ implies that for short-times it is always optimal to set 
$\|\mathfrak{a}\|=0$ in \eqref{QFI-growth}, as in this regime the linear growth dominates the quadratic one. This results in the QFI growing as $F_Q \approx 4 b_+^2 t^2$, where $b_+ = \min_{h, \|\mathfrak{a}\| = a_-}(\|\mathfrak{b}(h)\|)$.  When $a_- > 0$, however, the linear term cannot be removed and the QFI grows as $F_Q \approx  4 a_- t$. See Fig.~\ref{fig:abplot}(ii), where all the four cases of short-time and asymptotic scalings are depicted.

Furthermore, one can asses the characteristic time-scales of transition from short-time to asymptotic-time scaling, when contributions from the linear and quadratic terms become comparable. 
In the quadratic-linear case this leads to transition time $\tau \approx \frac{a_+}{b_+^2}$. In three other cases we in fact deal with two time scales---when the initial QFI formula ceases to be valid ($\tau_-$) and when the the QFI approaches the asymptotic behaviour ($\tau_+$). These times read $\tau_{\mp} \approx \frac{a_+}{b_{\pm}^2}$, $\tau_{\mp} \approx \frac{a_{\mp}}{b_{+}^2}$ and $\tau_\mp \approx \frac{a_\mp}{b_\pm^2}$ for quadratic-quadratic, linear-linear and linear-quadratic models (see \cite{supp}, Section B for details) respectively, see Fig.~\ref{fig:abplot}(ii).
%\wg{[see footnote~\footnote{More precisely, these times are valid assuming that the line joining point $(b_-,a_+)$ with $(b_+,a_-)$ in the `ab-plot' Fig. \ref{fig:abplot}(i) is far from dotted lines. Otherwise, transition times may be much longer}]}.

\paragraph*{Algebraic conditions.} While obtaining the exact values of $a_\pm$, $b_\pm$ requires in general to run an SDP programme, determining whether $a_-$, $b_-$ are zero or strictly positive may be phrased in terms of simple algebraic conditions on $H$, $\dot{H}$, $\vec{L}$ and $\dot{\vec{L}}$.
Inspecting \eqref{eq:aoperator}, we see that the condition for the \emph{quadratic short-time scaling} can be phrased as:
\begin{equation}
\label{eq:a0cond}
a_- = 0 \quad \t{iff} \quad\exists_{\mathfrak{h},\vec{h}} \, \dot{\vec{L}} =-i(\mathfrak{h} \vec{L} + \vec{h}\openone ).
\end{equation}
Looking at \eqref{eq:boperator} it is clear that the condition for \emph{linear asymptotic-time scaling} amounts to
\begin{equation}
\label{eq:b0cond}
b_-=0 \quad \t{iff} \quad G = \dot{H} - \frac{i}{2} \left( \dot{\vec{L}}^\dagger \vec{L} - \vec{L}^\dagger \dot{\vec{L}} \right) \in \mathcal{S},
\end{equation}
where $\mathcal{S}=\t{span}_{\mathbb C}\{\openone, L_k,L_k^\dagger, L_k^\dagger L_l\}$ has been called the Lindblad-span space in \cite{Demkowicz-Dobrzanski2017, Zhou2017}.

If the parameter is encoded in $H$ only, so $\dot{\vec{L}}=0$, the condition \eqref{eq:a0cond} is trivially satisfied and the initial scaling of QFI is always quadratic.  The impact of parameter dependence of $L_k$ operators on the asymptotic scaling has also been analyzed in \cite{chen2020, Wan2022}, but a general characterization of short and asymptotic scaling conditions was not given.

To investigate the impact of non-zero $\dot{\vec{L}}$ consider the first the case when $\dot L_i\in\t{span}_{\mathbb C}\{\openone,L_k\}$. This allows us to write them as
\begin{equation}
\label{eq:dLvec}
    \forall_i\dot L_i\in\t{span}_{\mathbb C}\{\openone,L_k\}\Leftrightarrow \dot{\vec{L}}=\mathfrak{h}_1\vec{L}-i\mathfrak{h}_2\vec{L}-i\vec{h}\openone,
\end{equation}
where $\mathfrak{h}_1$ is hermitian matrix, $-i\mathfrak{h}_2$ is anti-hermitian matrix, and $-i\vec{h}$ is a complex vector. 
Thanks to the freedom of redefining $L_k$ and $H$, while keeping the quantum master equation unchanged, one can show that there is a representation $H^\prime$, $L_k^\prime$ where the part $-i\mathfrak{h}_2$ disappears, while $-i\vec{h}\openone$ part is transferred to the Hamiltonian (see Ref. \cite{supp}, Section C):
\begin{equation}
\dot{\vec{L}}^\prime =  \mathfrak{h}_1\vec{L}, \quad \dot{H}^\prime= \dot{H} - \frac{1}{2}
\left[\vec{h}^\dagger \vec{L} + \vec{L}^\dagger \vec{h} \right].
\end{equation}
Hence, only the term $\mathfrak{h}_1\vec{L}$ is responsible for the parameter dependence truly connected with the actual dissipative part of the evolution. If this part is non zero, \eqref{eq:a0cond} cannot be satisfied. Consequently, we can always interpret the initial quadratic scaling of QFI with a \emph{purely} Hamiltonian parameter estimation model, while the appearance of linear initial scaling is a clear indication of non-trivial parameter dependence of noise operators $L_k$. Moreover, the effective modification of Hamiltonian due to $-i\vec{h}\openone$ term, will not affect the \eqref{eq:b0cond} condition as the added term lies in $\mathcal S$. Hence, on its own it will never allow for a Heisenberg scaling asymptotic behaviour.

%In contrary, if $\mathfrak{h}_1= 0,i\mathfrak{h}_2\neq 0$, the condition \eqref{eq:a0cond} may be trivially satisfied. Moreover, even if $- \frac{i}{2} \left( \dot{\vec{L}}^\dagger \vec{L} - \vec{L}^\dagger \dot{\vec{L}} \right)\neq0$ here, it still belong to $\mathcal S$. Therefore, even if it may have an impact for finite times, the condition \eqref{eq:b0cond} is not affected by the dependence of jump operators on the estimated parameter, and the asymptotic linear scaling is determined by the standard condition $\dot{H} \in \mathcal{S}$. 

%To sum up, $\mathfrak{h}_1$ is responsible for the parameter non-trivially encoded in the noise operators, which also implies linear scaling for all times. The contrary $i\mathfrak{h}_2$ part may be effectively understood as the changes of Hamiltonian, which, however, still does not allow for HS.

Finally, the interesting case is when at least one of the derivatives $\dot{L}_i$ does not live in the span of Lindblad operators $\exists_i \dot L_i\notin\t{span}_{\mathbb C}\{\openone,L_k\}$. Then the second term in \eqref{eq:b0cond} provides a nontrivial contribution. In particular, we may even observe asymptotic quadratic QFI scaling, even with no Hamiltonian parameter dependence at all ($\dot{H}=0$) or when the Hamiltonian part on its own would not allow for this--see Examples below. 

\begin{figure*}[t]
\includegraphics[width=1.7\columnwidth]{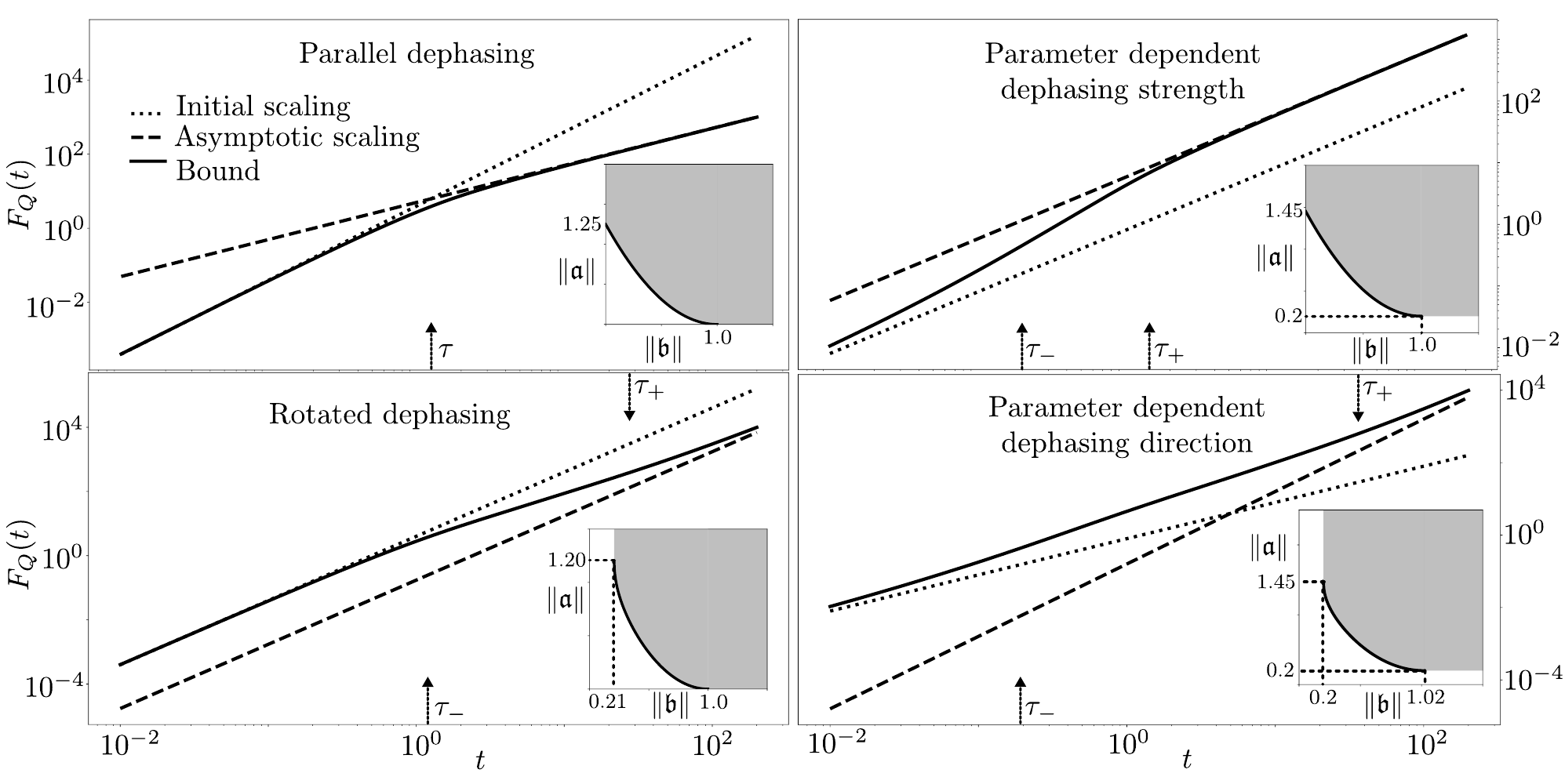}
\caption{Bound on $F_Q(t)$ derived using \eqref{formula-QFI}, as well as initial and asymptotic scalings of $F_Q(t)$ as a function of time $t$ in log-log scale for four different sensing cases in the same order as depicted in Fig.~\ref{fig:abplot}(ii). $\tau$, $\tau_-$, and $\tau_+$ denote the transition times computed according to formulas presented in the schematic Fig. \ref{fig:abplot}(ii).
 The insets represent the corresponding `ab-plots' indicating the values of $a_-$, $b_-$, $a_+$, and $b_+$. The structure and the numerical parameters for the models are given in Table~\ref{tab:models}.}
\label{fig:examples}
\end{figure*}

\paragraph*{Quantitative analysis.} We now present a recipe to obtain the tightest possible numerical bounds within the framework presented. The method amounts to a  discrete time step approximated integration of \eqref{QFI-growth}, during which $\|\mathfrak{a}(h)\|$, $\|\mathfrak{b}(h)\|$ are chosen at each step in a way to yield the tightest bound possible. Note that, 
since we want our bound to be fundamentally valid, we \emph{do not} assume that control operations can be applied only on time scales longer than the discretization time step $\Delta t$.
%---in this case we would trivially go back to the channel estimation framework discussed in \cite{adaptive-new-bound}. 
On the contrary, our bound will be valid for arbitrary fast and arbitrary strong quantum controls, while the choice of the $\Delta t$ will only affect the tightness of the bound---the smaller $\Delta t$ the tighter the bound. 

We first note, that RHS of \eqref{QFI-growth} is strictly increasing with time, therefore for any $h$, $F_Q(t+\Delta t)-F_Q(t)\leq 4(\mathfrak{a}(h)+\mathfrak{b}(h)\sqrt{F_Q(t+\Delta t)})\Delta t$. Next, thinking of $F_Q(t+\Delta t)$ and $F_Q(t)$ as independent variables, we solve the above for $F_Q(t+\Delta t)$. Finally, as the obtained inequality is valid for any $h$, we may optimize it over $h$:
%\begin{multline}
%F(t+\Delta t)\leq F(t)+4\min_h \Big[\mathfrak{a}(h)\Delta t+8\mathfrak{b}(h)\Delta t^2+\\
%\mathfrak{b}(h)\Delta t\sqrt{F(t)+4\mathfrak{a}(h)\Delta t^{3/2}+4\mathfrak{b}(h)^2\Delta t^2}\Big].
%\end{multline}
%\wg{corrected [to be double checked again]:
\begin{multline}
\label{formula-QFI}
F_Q(t+\Delta t)\leq F_Q(t)+4\min_h \Big[\|\mathfrak{a}(h)\|\Delta t+2\|\mathfrak{b}(h)\|^2\Delta t^2+\\
\|\mathfrak{b}(h)\|\Delta t\sqrt{F_Q(t)+4\|\mathfrak{a}(h)\|\Delta t+4\|\mathfrak{b}(h)\|^2\Delta t^2}\Big].
\end{multline}
Clearly, in the limit $\Delta t \rightarrow 0$ the above formula corresponds to exact integration of $\eqref{QFI-growth}$, but the important point is that for any finite $\Delta t$ it results in a fundamentally valid bound. It is not advisable to optimize $h$ at every step via an SDP. It is much more efficient to first obtain the `ab-plot' and then at each step identify the point on the curve $a(b)$ that provides the tightest bound at a given step---this can be done by a simple search through tabularized values of $a(b)$.

\paragraph*{Examples.}
In Fig.~\ref{fig:examples} we present numerical results for four models representing all four different `scaling cases', ordered in a way to correspond to classification presented in Fig.~\ref{fig:abplot}(ii).    
The results are presented in log-log scale in order to better reveal the scaling character of the curves. In all the models considered the Hamiltonian is $H(\varphi)=\omega \varphi\sigma_z$, and they differ only by the form of the single noise operator $L_1(\varphi)$, which is given together with numerical parameters of the models is Table~\ref{tab:models}. 
Without loss of generality, we set $\omega=1$ which therefore determines the natural time scale for the problem to be $\omega^{-1}$.
We also assume that the parameter to be estimated $\varphi$ is dimensionless, as a result the QFI will also be dimensionless.

\begin{table}[t]
    \begin{tabular}{c||c|c|c|c|c|c|c}
    model & $L_1(\varphi)$ & $a_-$ & $b_-$ & $a_+$ & $b_+$ & $\tau_-$ & $\tau_+$\\
    \hline\hline
       PD & $\sqrt{\frac{\gamma}{2}}\sigma_z$  &  $\boldsymbol{0}$ & $\boldsymbol{0}$ & $1.25$  & $1$ &  \multicolumn{2}{c}{$1.25$}  \\
       \hline
       RD  & $\sqrt{\frac{\gamma}{2}} \sigma\!\left(\frac{\pi}{15}\right)$  & $\boldsymbol{0}$  & $0.21$ & $1.20$  & $1$ & 1.20 & 27.67 \\
       \hline
       PDDS & $\sqrt{\frac{\gamma}{2}} e^\varphi \sigma_z$  & $0.2$  & $\boldsymbol{0}$ &  $1.45$ & 1 & 0.2 & 1.45 \\
       \hline
       PDDD &  $\sqrt{\frac{\gamma}{2}}\sigma \!\left( \varphi \right)$ & 0.2  & 0.2 & 1.45  & 1.02  & 0.19 & 36.25\\
    \end{tabular}
    \caption{The form of the single noise operator $L_1$ and the parameters of the models depicted in Fig.~\ref{fig:examples}: PD (parallel dephasing), RD (rotated dephasing), PDDS (parameter dependent dephasing strength), PDDD (parameter dependent dephasing direction), where 
    $\sigma( \theta) = \sigma_z \cos \theta +\sigma_x \sin \theta$.
    In all the models $H(\varphi)= \varphi \omega \sigma_z$. Numerical values are given for the parameters set to $\omega =1$, $\gamma=0.4$. The sensing is performed around the parameter value $\varphi=0$.} 
    \label{tab:models}
\end{table}

We take the conventional parallel dephasing (PD) model as an example of quadratic-linear case ($a_- =0 $, $b_-=0$). This may be regarded as a generic Hamiltonian parameter estimation case, where noise cannot be removed so that asymptotic Heisenberg scaling is recovered. Time $\tau$ shown in the plot is indeed a good indicator of initial quadratic to asymptotic linear scaling transition. 
 
For quadratic-quadratic case, we consider a rotated dephasing  (RD) model where $G=\dot{H} \propto \sigma_z\notin \mathcal{S}$ and $\dot{L_1}=0$, resulting $a_-=0$, but $b_->0$. We have two transition times--- $\tau_-$ indicating the transition away from initial quadratic scaling and $\tau_+$ indicating the transition to asymptotic quadratic scaling.

For the linear-linear case, we consider a the parameter dependent dephasing strength (PDDS) model, such that $a_- > 0$, but $b_-=0$. Two transition times $\tau_-$, and $\tau_+$ faithfully indicates the transition from initial linear scaling to intermediate scaling and then transition to asymptotic linear scaling respectively. In Ref. \cite{Sekatski2022}, non-equilibrium quantum thermometry for Markovian master equation is considered, where the jump operators are temperature dependent. This is an example of linear-linear case.

Finally, we consider a parameter dependent dephasing direction (PDDD) model, where $a_{-} > 0$, and $b_{-} > 0$. This model has also been studied in \cite{Wan2022} being an example where asymptotic Heisenberg scaling is caused by the parameter dependent jump operator. The two time scales shown in the plot are $\tau_-$, and $\tau_+$, denoting the transition from initial linear scaling to intermediate, and ultimately to asymptotic quadratic scaling respectively. 

In some situation it is possible to derive an analytical form of the bound obtained via numerical optimization. We provide such an example in Ref. \cite{supp}, Section D for the PD case, but stress that in general SDP optimization is necessary to obtain the tightest bound.  

\paragraph*{Conclusions.}
With this paper, we performed comprehensive analysis of general Markovian sensing models and provided simple and efficient tools to assess the potential quantum enhancement for the models on all evolution
time-scales. Notice that the framework covers in particular all continuous-measurement paradigms \cite{Rossi2020, Binefa2021} and as such any results obtained in these approaches should obey the bounds derived here.
An obvious direction for further research is to study non-Markovian models \cite{Chin2012, Smirne2016}, but this would require going significantly beyond the framework presented here.

\paragraph*{Note added.} 
Recently, Ref. \cite{ChenPRL2024} appeared, where authors derived a condition analogous to (\ref{eq:b0cond}), providing a complementary approach to the similar context.

\begin{acknowledgements}
We thank Francesco Albarelli and Staszek Kurdzia{\l}ek  for fruitful discussions.
This work was supported by National Science Center (Poland) grant
No.2020/37/B/ST2/02134.
WG acknowledges support from the U.S. Department of Energy, Office of Science, National Quantum Information Science Research Centers, Superconducting Quantum Materials and Systems Center (SQMS) under contract number DE-AC02-07CH11359.
\end{acknowledgements}

\bibliography{timescales}

\appendix

 \section{Semi-definite programme to generate the $(\|\mathfrak{a}\|,\|\mathfrak{b}\|)$-plot}
\label{sec:abplot}

To generate the `ab-plot' plot, first we determine ${a}_{-} = \min_h\| \mathfrak{a}(h)\|$ and 
${b}_- = \min_h \|\mathfrak{b}(h)\|$ according to the main text. Next step is to calculate ${b}_+ = \min_{h, \|\mathfrak{a}\| = {a}_-}(\|\mathfrak{b}(h)\|)$ and then evaluating $\min_h \|\mathfrak{a}\|$, for a given $b\in [b_-,b_+]$. Clearly, given $b=b_-$, we get $\min_h \|\mathfrak{a}\|=a_+$. To be able to do the above calculations, we formulate the following optimization problems,
\begin{align}
\label{formulate-SDP-1}
    & {\rm min}_{h} \|\mathfrak{a}\|,~~\text{subject to} ~~ \|\mathfrak{b}\|\leq b,\\
    \label{formulate-sdp-2}
    & {\rm min}_{h} \|\mathfrak{b}\|,~~\text{subject to} ~~ \|\mathfrak{a}\|\leq a.
\end{align}
Note that the above minimization problems simply produce $a_-$, and $b_-$ without any constraints on $\|\mathfrak{b}\|$, and $\|\mathfrak{a}\|$ respectively.
We can pose the above optimizations as SDP by constructing the following matrices,
\begin{align}
   A=
    &\begin{pmatrix}
        a\mathds{1} &{\left(i\dot{\vec{L}}+ \mathfrak{h} \vec{L} + \vec{h} \openone \right)}^\dagger\\
        {\left(i\dot{\vec{L}}+ \mathfrak{h} \vec{L} + \vec{h} \openone \right)} & \mathds{1}^{\otimes n}
    \end{pmatrix},\\
    & B=
    \begin{pmatrix}
        b\mathds{1} & i\mathfrak{b}(h)\\
        -i\mathfrak{b}^\dagger(h) & b\mathds{1}
    \end{pmatrix}.
\end{align}
Using the Schur's complement condition, one can show that, $A\succeq 0 \iff	a\geq \|\mathfrak{a}\|$, and similarly $B\succeq 0 \iff	b\geq \|\mathfrak{b}\|$. As a result, just setting the constraint as $A\succeq 0$, and minimizing $a$ is equivalent to minimizing $\|\mathfrak{a}\|$, which gives us $a_-$. Similarly, minimizing $b$ with the constraint $B\succeq 0$ is equivalent to minimizing $\|\mathfrak{b}\|$, producing $b_-$. Furthermore, minimizing $a$, with two constraints $A\succeq 0$, and $B\succeq 0$, for a given $b$ leads to the solution of the optimization problem defined in Eq. (\ref{formulate-SDP-1}). 
Likewise, minimizing $b$, with the constraints $A\succeq 0$, and $B\succeq 0$ for a given $a$ produces solution of Eq. (\ref{formulate-sdp-2}). Clearly, setting $a=a_-$, we obtain $b_+$. Then we solve Eq. (\ref{formulate-SDP-1}), setting $b\in [b_-,b_+]$, and thus producing the `ab-plot'.

\section{Derivation of transition times}
To asses the characteristic timescales when the transition from the short-time to the asymptotic-time scaling regime takes place, it is enough to compare the contributions from the linear and quadratic terms, and find time $\tau$ when they become comparable. For quadratic-linear case  ($a_-=b_-=0$), initially, making linear contribution of QFI zero gives the tightest bound as the linear growth dominates the quadratic one. Consequently, QFI grows as $F_Q\approx 4b_+^2t^2$ for short-times. However, gradually quadratic contribution starts to dominate and after time $\tau$, tightest bound on $F_Q$ is obtained by making $\|\mathfrak{b}\|=0$, resulting linear growth of $F_Q\approx 4a_+t$, which is also the asymptotic scaling. So, the characteristic transition time $\tau$ between quadratic and linear behavior is obtained when $4b_+^2\tau^2\approx 4a_+\tau$. From this we get, $\tau\approx \frac{a_+}{b_+^2}$. For other three cases we actually get two characteristic time scales---when the initial QFI formula ceases to be valid ($\tau_-$) and when the the QFI approaches the asymptotic behavior ($\tau_+$).

For quadratic-quadratic case ($a_-=0$, and $b_->0$), just like the previous scenario, initially QFI grows quadratically---$F_Q\approx 4b_+^2t^2$. After a time $\tau_-$, when quadratic contribution starts to dominate over linear term, setting $\|\mathfrak{a}\|=0$ no longer gives the optimal bound. Instead, optimal bound will be obtained by suppressing the quadratic contribution as much as possible. So, we will see departure from the initial quadratic scaling when it becomes comparable to the linear part corresponding to $\|\mathfrak{b}\|=b_-$: $4b_+^2\tau_-^2\approx 4a_+\tau_-$.
and we get the first transition time to be $\tau_-\approx \frac{a_+}{b_+^2}$. Finally, in the asymptotic limit, the QFI scaling is necessarily quadratic--- $F_Q\approx 4b_-^2 t^2$ as the linear contribution can be neglected. Hence, the second transition timescale from intermediate scaling to the asymptotic quadratic scaling is obtained from $4a_+\tau_+\approx 4b_-^2\tau_+^2$, giving $\tau_+\approx \frac{a_+}{b_-^2}$. 

For linear-linear case ($a_->0$, and $b_-=0$), as $\|\mathfrak{a}\|$ can not be zero, we can not remove the linear contribution. As a result, at very short-times, QFI growth is essentially linear--- $F_Q\approx 4a_-t$. However, after a time $\tau_-$, quadratic contribution can not be neglected anymore and we will see the departure from initial linear scaling when $4a_-\tau_-\approx 4b_+^2\tau_-^2$. This gives the first transition time $\tau_-\approx \frac{a_-}{b_+^2}$. In the asymptotic limit of-course the tightest bound is obtained by setting $\|\mathfrak{b}\|=0$ and we get $F_Q\approx 4a_+ t$. So, the second transition time is $\tau_+\approx \frac{a_+}{b_+^2}$, after which quadratic term starts to dominate over the linear contribution.

Finally, for the linear-quadratic case ($a_->0$, and $b_->0$), the initial scaling is again linear given by $F_Q\approx 4a_-t$. Just like the linear-linear case, the first transition time $\tau_-\approx \frac{a_-}{b_+^2}$. Asymptotically, the QFI scaling is quadratic given as $F_Q\approx 4b_-^2t^2$. This scaling is obtained after a time $\tau_+$, when the quadratic scaling starts to dominate over the linear scaling: $4a_+\tau_+\approx 4b_-^2\tau_+^2$. Hence, the second transition time is $\tau_+\approx \frac{a_+}{b_-^2}$.

\section{The class of models with effectively only Hamiltonian parameter dependence}
\label{sec:Honly}

Let us assume that the dynamics of the sensing system is governed by the quantum master equation \eqref{GKSL} with Hamilotnian $H$ that has an arbitrary dependence on estimated parameter $\varphi$, and jump operators $\vec{L}$ such that $\dot{\vec{L}}=\mathfrak{h}_1\vec{L}-i\mathfrak{h}_2\vec{L}-i\vec{h}\openone$. Quantum master equation is invariant under a unitary mixing of jump operators (homogeneous transformation) as well as the inhomogeneous transformation mixing the Hamiltonian with the jump operators \cite{breuer02}:  
\begin{equation}
\begin{split}
 \vec{L}^\prime &=   u(\varphi)\left[\vec{L} + \vec{v}(\varphi) \openone\right],\\
 H^\prime &= H + 
 \frac{1}{2i}\left[\vec{v}^\dagger(\varphi) \vec{L} - \vec{L}^\dagger \vec{v}(\varphi)\right],
 \end{split}
\end{equation}
where $u(\varphi)$ is a $n \times n$ unitary matrix and  $\vec{v}(\varphi) \in \mathbb{C}^n$. 
We choose  $u(\varphi) = e^{i \mathfrak{h}_2 \varphi}$ (matrix exponentiation) and $\vec{v}(\varphi) = e^{i \vec{h} \varphi} - \vec{1}$, where we perform element-wise exponentiation and where $\vec{1}$ denotes a vector of length $n$ with all entries equal to $1$. Finally, assuming the derivatives are taken at $\varphi=0$ (this is no loss of generlity as we can always shift the value of parameter in the definition), we get:
\begin{equation}
\dot{\vec{L}}^\prime =  \mathfrak{h}_1 \vec{L}, \quad \dot{H}^\prime= \dot{H} - \frac{1}{2}
\left[\vec{h}^\dagger \vec{L} + \vec{L}^\dagger \vec{h} \right].
\end{equation}
 As a result, we obtain an equivalent model, where the parameter enters into the Lindblad operator only via hermitian part $\mathfrak{h}_1$. If $\mathfrak{h}_1=0$, 
the parameter enters into the Hamiltonian part of the master equation only.

 \section{Analytical approximately optimal formula for the bounds}
 \label{app:analitical}

In general, we obtain the tightest bound only numerically. Yet, some useful sub-optimal bounds may be easily derived, following the reasoning from \cite{Wan2022}, adapted to our formalism.

For any $h$ (fixed to be the same during the whole evolution), the inequality:
\begin{equation}
\frac{dF(t)}{dt}\leq 4\|\mathfrak{a}(h)\|+4\|\mathfrak{b}(h)\|\sqrt{F(t)}
\end{equation}
may be directly integrated (to a rather complicated form) and further bounded by (Eq. (C17) of \cite{Wan2022}):
\begin{equation}
\label{eq:simplegen}
    F(t)\leq 4\left(\|\mathfrak{b}(h)\|t+\sqrt{\|\mathfrak{a}(h)\|t}\right)^2.
\end{equation}
The above bound is valid for any model. In limits $t\to 0$ and $t\to\infty$ it gives respectively $\sim 4\|\mathfrak{a}(h)\|t$ and $\sim 4\|\mathfrak{b}(h)\|^2t^2$.

For the quadratic-linear case ($a_-,b_-=0$) an analytical bound tighter than above may be derived. Let us name by $h_a$, $h_b$ a proper $h$ matrices satisfying:
\begin{alignat}{2}
&\|\mathfrak{a}(h_b)\|=0,\quad &&\|\mathfrak{b}(h_b)\|=b_+,\\
    &\|\mathfrak{a}(h_a)\|=a_+,\quad &&\|\mathfrak{b}(h_a)\|=0.
\end{alignat}
Then for linear combination $h=\delta h_a+(1-\delta)h_b$, where $0\leq \delta\leq 1$, we have:
\begin{multline}
    \left(i\dot{\vec{L}}+ (\delta\mathfrak{h}_a+(1-\delta)\mathfrak{h}_b \vec{L} + (\delta\vec{h}_a+(1-\delta)\vec{h}_b) \openone\right)=\\
    \delta \left(i\dot{\vec{L}}+ \mathfrak{h}_a \vec{L} + \vec{h}_a \openone\right)+\underbrace{(1-\delta)\left(i\dot{\vec{L}}+ \mathfrak{h}_b \vec{L} + \vec{h}_b \openone\right)}_{=0}
\end{multline}
so from Eqs. \eqref{eq:aoperator} and \eqref{eq:boperator} of main text,
\begin{alignat}{2}
\label{a-gen}
&\|\mathfrak{a}(\delta h_a+(1-\delta)h_b)\| =\quad\delta^2\|\mathfrak{a}(h_a)\|&&=\quad\delta^2a_+\\
\label{b-gen}
&\|\mathfrak{b}(\delta h_a+(1-\delta)h_b)\|=(1-\delta)\|\mathfrak{b}(h_b)\|&&=(1-\delta)b_+.
\end{alignat}
Then Eq. \eqref{QFI-growth} of main text takes the form:
\begin{equation}
    \frac{dF(t)}{dt}\leq 4\delta^2 a_++4(1-\delta)b_+\sqrt{F(t)}.
\end{equation}
RHS may be directly minimized by the choice of $\delta$:
\begin{equation}
    \delta=\begin{cases}
        \frac{b_+}{2a_+}\sqrt{F(t)}\quad &\t{for}\, \frac{2a_+}{b_+}\leq \sqrt{F(t)},\\
        1\quad &\t{for}\, \frac{2a_+}{b_+}\geq \sqrt{F(t)}.
    \end{cases}
\end{equation}
Then the bound for $F(t)$ may be integrated directly:
\begin{equation}
\label{eq:simpleQL}
    F(t)\leq\begin{cases}
       \frac{16a_+^2}{b_+^2}(1-2^{-t/t_c})^2\quad &\t{for}\, t\leq t_c,\\
        4a_+(t-t_c)+4(\frac{a_+}{b_+})^2\quad &\t{for}\, t\geq t_c,
    \end{cases}
\end{equation}
where $t_c=\frac{2a_+}{b_+^2}\ln 2$ (for comparison see Eq. (31) of \cite{Wan2022}, which coincides with our bound after the substituations $a_+\leftrightarrow c_2$, and $b_+\leftrightarrow c_1$). 

For the PD model considered above, the bound \eqref{eq:simpleQL} turns out to be as tight as the one obtained by full numerical optimization. This is, however, not the case for all quadratic-linear models, and in principle, for some $t$ the optimal $h(t)$ may be not the one of the form $\delta h_a+(1-\delta)h_b$.

Similar construction of the bound can be carried out for quadratic-quadratic case leading to the analogous bound as  Eq. (37) of \cite{Wan2022} with the substitutions $a_+\leftrightarrow c_2$, $b_+\leftrightarrow c_1$, and $b_-\leftrightarrow c_0$. But this bound is not optimal for the RD model in the main text as optimal $h(t)$ is not of the form $\delta h_a+(1-\delta)h_b$. 

Moreover, for linear-linear and linear-quadratic, this method of derivation does not work as there are no counterparts of Eq.  (\ref{a-gen}) and (\ref{b-gen}) solely in terms of the constants $a_+$, $b_+$, $a_-$, and $b_-$. For that we need to use Eq. (\ref{formula-QFI}) of main text to obtain the tightest numerical bound, while there is no simple recipe to provide the tightest bound from Eq. (18) in the Ref \cite{Wan2022}. 

%\bibliography{timescales}

\end{document}